\documentclass[aps,amsmath,amssymb,twocolumn,showpacs]{revtex4-1}
\usepackage{bm}
\usepackage{graphicx}
\usepackage{color}  
\usepackage{epstopdf}
\usepackage{amsmath}
\usepackage{ulem}
\usepackage[unicode=true,colorlinks=true,citecolor=blue]{hyperref}
\begin{document}
\title{Valley polarization induced second harmonic generation in graphene}
\author{L.\,E.\,Golub$^1$}
\author{S.\,A.\,Tarasenko$^{1,2}$}
%
\affiliation{$^1$Ioffe Physical-Technical Institute, 194021 St.\,Petersburg, Russia}
\affiliation{$^2$St.\,Petersburg State Polytechnic University, 195251, St.\,Petersburg, Russia}
\begin{abstract}
The valley degeneracy of electron states in graphene stimulates intensive research of valley-related optical and transport phenomena. 
While many proposals on how to manipulate valley states have been put forward, experimental access to the valley polarization in graphene is still a challenge. Here, we develop a theory of the second optical harmonic generation in graphene and show that this effect can be used to measure the degree and sign of the valley polarization. We show that, at the normal incidence of radiation, the second harmonic generation stems from imbalance of carrier populations in the valleys. The effect  has a specific polarization dependence reflecting the trigonal symmetry of electron valley and is resonantly enhanced if the energy of incident photons is close to the Fermi energy.
\end{abstract}

\maketitle


The valley degree of freedom of charge carriers in multi-valley semiconductor systems such as silicon, diamond, graphene, carbon nanotubes, transition metal dichalcogenides, etc. attracts growing attention due to great and yet unexplored potential of semiconductor valley properties to practical applications~\cite{Rohling2012,Laird2013}. A promising candidate for the study of valley physics in two dimensions is graphene, a one-atom-thick layer of carbon~\cite{CastroNeto}. Graphene technology is now well developed, which enables the synthesis of large-scale defect-free monolayers as well as the production of graphene nanostructures with controllable shapes and edges~\cite{Yu2011,Murdock2013,Xu2013}. A number of proposals on how to generate the valley polarization of carriers and valley currents in graphene has been put forward. It was shown that the electric current gets valley polarized in a graphene point contact with zigzag edges~\cite{Rycerz07}, graphene layer with broken inversion symmetry~\cite{Xiao07}, strained graphene with mass Dirac fermions~\cite{Grujic2014}, at the boundary between monolayer and bilayer graphene~\cite{Nakanishi10,Pratley2014}, at a line defect~\cite{Gunlycke11}, or if monolayer or bilayer graphene is additionally illuminated by circularly polarized radiation~\cite{Oka09,Abergel09}. Valley currents can be induced in graphene rings by asymmetric monocycle electromagnetic pulses~\cite{Moskalenko09}. It was also proposed that bulk valley currents in graphene and carbon nanotubes can be excited by polarized light~\cite{Golub2011,TarIvch05,Hartmann2011} or ac mechanical vibrations~\cite{Jiang2013}. While the above methods can be used to create imbalance in valley populations, experimental study of valley phenomena and verification of the theoretical proposals is still a challenge because of lack of efficient and reliable methods to probe the valley polarization.

The valleys in graphene are situated at the $K$ and $K'$ points of the two-dimensional Brillouin zone which are connected with each other by the space inversion $C_i$~\cite{McCann06}. Each of the valleys is described by the $D_{3h}$ small group and lacks the center of space inversion while the overall symmetry of free standing graphene $D_{6h} = D_{3h} \times C_i$ is centrosymmetric. It follows that the valley polarization of carriers reduces the spatial symmetry of the structure to the symmetry of an individual valley. Such a symmetry reduction gives rise to optical effects such as second harmonic generation (SHG) which require the spatial symmetry breaking~\cite{Golub2011}. Here, we develop a microscopic theory of SHG in graphene and show that the effect can be used to measure the degree and the sign of valley polarization. We demonstrate that valley polarization induced SHG is caused by the trigonal warping of the electron dispersion in valleys and calculate the second-order susceptibility tensor for interband optical transitions. The efficiency of SHG is resonantly enhanced if the energy of incident photons is close to the Fermi energy. The second optical harmonic due to the valley polarization is generated at the normal incidence of radiation and, therefore, can be discriminated from the SHG signals stemming from structure inversion asymmetry of graphene flakes on substrate~\cite{Dean2010} or in-plane photon momentum~\cite{Mikhailov2008,Glazov2011} which both require the oblique incidence of radiation. The effect is also different by symmetry from SHG caused by the flow of a direct electric current in the sample~\cite{Wu2012,Bykov2012,Cheng2014}. Since non-linear optical spectroscopy is a sensitive and powerful tool to study carrier kinetics and structure symmetry with high spatial resolution, SHG applied to graphene will enable the local probe of valley polarization as well as the study of valley polarization thermal fluctuations (valley noise)~\cite{Tse2014}. 

\begin{figure}[t]
\includegraphics[width=0.8\linewidth]{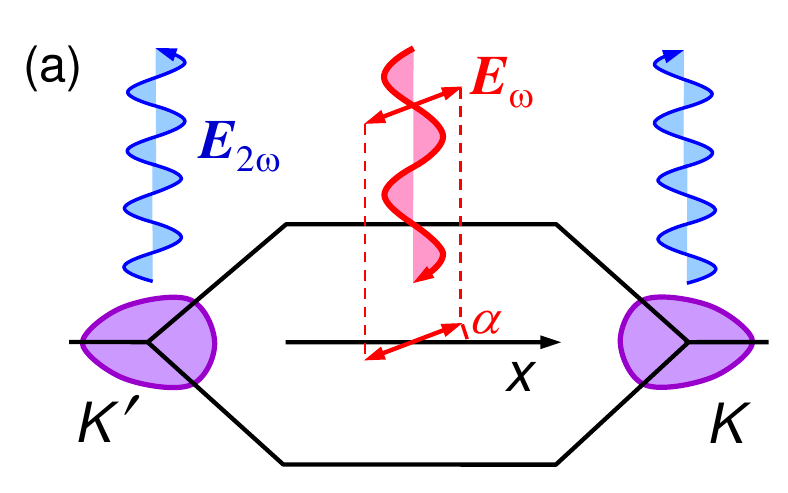}\\ \vspace{0.2cm}
\includegraphics[width=0.8\linewidth]{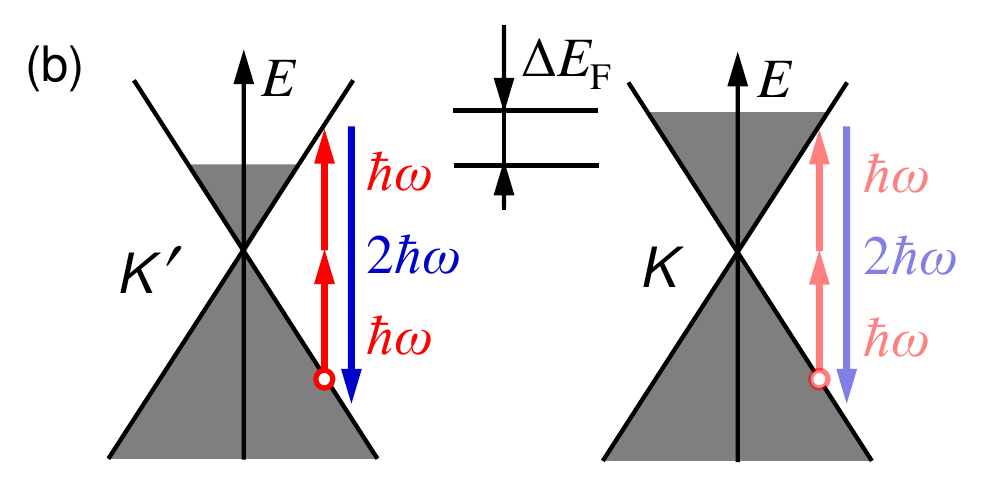}
\caption{Microscopic mechanism of valley polarization induced second harmonic generation. (a) SHG is caused by trigonal asymmetry of the valleys. Signals stemming from the $K$ and $K'$ valleys have opposite sign. (b) Imbalance of valley populations gives rise to a net SHG signal. SHG is resonantly enhanced if the energy of incident photons is close to the Fermi energy.}
\label{fig_SHG}
\end{figure}

The valley polarization induced SHG is illustrated in Fig.~\ref{fig_SHG}. We assume that graphene is excited by a plane electromagnetic wave with the frequency $\omega$ at the normal incidence. Both $K$ and $K'$ valleys have trigonal symmetry and contribute to SHG. However, since the valleys are related to each other by the space inversion, SHG signals stemming from the valleys are counter phased,  Fig.~\ref{fig_SHG}a. Therefore, the total SHG signal vanishes at equal distribution of carriers in the valleys and arises in the case of imbalance of the valley populations, see Fig.~\ref{fig_SHG}b. SHG has a resonant behavior and drastically enhanced if the energy of incident photon $\hbar\omega$ is close to the Fermi energy of carriers, Fig.~\ref{fig_SHG}b.  

Phenomenologically, SHG is described by the second-order susceptibility tensor $\bm{\chi}$ which couples the polarization amplitude at the double frequency $\bm{P}(2\omega)$ with the incident radiation electric field amplitude $\bm{E}(\omega)$,
\begin{equation}
	P_\alpha (2\omega)= \chi_{\alpha\beta\gamma} E_\beta(\omega) E_\gamma(\omega) \:,
\end{equation}
where $\alpha$, $\beta$, and $\gamma$ are the Cartesian coordinates. Here, we assume that the incident radiation is linearly polarized and its electric field has the form $\bm E(t) = \bm E(\omega) \exp{(-{\rm i}\omega t)} + \bm E (\omega) \exp{({\rm i}\omega t)} = 2\bm E(\omega) \cos{\omega t}$. Symmetry analysis shows that non-zero components of the tensor $\bm{\chi}$ caused by valley polarization are 
\begin{equation}\label{chi_comp}
	\chi_{xxx}=-\chi_{xyy}=-\chi_{yxy} \equiv \chi \:,
\end{equation}
where $x$ and $y$ are the in-plane axes perpendicular to each other with $x$ parallel to the $K'K$ direction, Fig.~\ref{fig_SHG}a. 
Equation~\eqref{chi_comp} implies the polarization dependence 
\begin{equation}\label{polarization}
	P_x(2\omega) = \chi|E(\omega)|^2 \cos{2\alpha} ,\, {P_y(2\omega) = -\chi|E(\omega)|^2\sin{2\alpha}} ,
\end{equation}
where $\alpha$ is an angle between the polarization plane of incident radiation and the $x$ axis, Fig.~\ref{fig_SHG}a.
Such a polarization behavior of SHG follows from the trigonal symmetry of an individual valley. 
The generation of second harmonic at the normal incidence of radiation and its specific polarization dependence given by
Eq.~\eqref{polarization} enable one to discriminate the valley-related SHG from other sources of SHG signal stemming, e.g., from structure inversion asymmetry of graphene flakes.

Microscopic calculation of the second-order susceptibility can be carried out in the density-matrix-theory formalism~\cite{Khurgin95}. In this approach, electron system in each valley is described by the density matrix $\rho$ which satisfies the quantum kinetic equation
\begin{equation}\label{kin_gen}
\frac{\partial \rho}{\partial t} = - \frac{{\rm i}}{\hbar} [H + V, \rho] + {\rm St} \rho \:.
\end{equation}
Here, $H$ is the Hamiltonian in the absence of radiation, 
\begin{equation}
H  =	\left( \begin{array}{cc}
	0&\nu v_0 p_- - \mu p_+^2\\
	\nu v_0 p_+ - \mu p_-^2&0
	\end{array}
	\right) ,
\end{equation}
$\nu$ is the valley index ($\nu = \pm 1$ for the $K$ and $K'$ valleys, respectively), $v_0$ is the electron velocity, $p_{\pm}=p_x \pm {\rm i} p_y$, $\bm{p}$ is the electron momentum, $\mu$ is the constant of trigonal warping, $V$ is the operator of electron-photon interaction, 
\begin{equation}
V = - \frac{e}{c} \, \bm{v} \cdot \bm{A} + \frac{e^2}{2 c^2} \sum_{\alpha\beta} \frac{\partial v_{\alpha}}{\partial p_{\beta}} A_{\alpha} A_{\beta} \:,  
\end{equation}
$e$ is the electron charge, $c$ is the speed of light, $\bm{v} =  \partial H/ \partial \bm{p}$ is the velocity operator, $\bm{A}=-i(c/\omega)\bm{E}(\omega)$ is the vector potential amplitude, and ${\rm St} \rho$ is the collision integral describing relaxation processes.  

Solution of the kinetic Eq.~\eqref{kin_gen} can be expanded in the series of the electric field amplitude,
\begin{equation}
\rho = \rho^{(0)} + [\rho^{(1)} {\rm e}^{-{\rm i}\omega t} + {\rm c.c.}] + [\rho^{(2)} {\rm e}^{-{\rm 2i}\omega t} + {\rm c.c.}] + \ldots \:, 
\end{equation}
where $\rho^{(0)}$ is the equilibrium density matrix, $\rho^{(1)} \propto E$, and $\rho^{(2)} \propto E^2$. Second harmonic is determined by the term $\rho^{(2)}$. We consider optical transitions between the valence ($v$) and conduction ($c$) bands in $n$-doped graphene. Straightforward  calculations show that the interband and intraband components of the density matrix $\rho^{(2)}$ for a given valley and momentum have the form
\begin{equation}\label{rho}
\rho_{cv}^{(2)} = {(e/c)^2 (\bm A \cdot \bm v_{cv}) [\bm A \cdot (\bm v_{cc}-\bm v_{vv})] \over (2\hbar\omega - E_{cv} + i \gamma)(\hbar\omega - E_{cv} + i \gamma)} (f_v-f_c) 
\end{equation}
\[
\;\;\; + \frac{(e/c)^2 \sum\limits_{\alpha\beta} (\partial v_{\alpha}/\partial p_{\beta})_{cv} A_{\alpha} A_{\beta}}{2(2\hbar\omega - E_{cv} + i \gamma)} (f_v-f_c) \:,
\]
\[
\rho_{cc}^{(2)} = - \frac{(e/c)^2(\bm{A}\cdot\bm{v}_{cv})(\bm{A}\cdot\bm{v}_{vc})}{(\hbar\omega-E_{cv}+i\gamma)(\hbar\omega+E_{cv}+i\gamma)} (f_v - f_c) \:,
\]
where $v_{cv}=v_{vc}^*$, $v_{cc}$, and $v_{vv}$ are the interband and intraband matrix elements of the velocity operator in
the valley, $E_{cv}$ is the energy gap between the valence and conduction bands, 
$\gamma/\hbar$ is the decay rate of the interband component of the density matrix, $f_{v}$ and $f_{c}$ are the equilibrium electron distribution functions in the valence and conduction bands. Below we assume for simplicity that $\gamma$ is independent of energy. The component $\rho_{vc}^{(2)}$ can be obtained from $\rho_{cv}^{(2)}$ by the complex conjugation and the replacement $\omega \rightarrow -\omega$;  component $\rho_{vv}^{(2)}$ is equal to $-\rho_{cc}^{(2)}$.

Polarization at the double frequency can be expressed in terms of the current density at the double frequency and is given by
\begin{equation}\label{P}
	\bm P(2\omega) = (i/2\omega) \bm{j}_{2\omega} =  (ie/\omega) \sum_{\bm p, \nu} {\rm Tr} \left( \rho^{(2)} \bm v \right) \:,
\end{equation}
where the spin degeneracy is taken into account and summation is performed over the momentum and the valley index.
Calculation of Eq.~\eqref{P} shows that the second-order susceptibility $\chi$ is the sum of intravalley contributions,
\begin{equation}\label{chi}
	\chi = \chi_{+} + \chi_{-} \:, 
\end{equation}
where
\begin{equation}\label{chi_nu}
	\chi_{\nu} =  - {\rm i} \left({e \over \omega}\right)^3  \sum_{\bm p} \left[f_v^{(\nu)}(-\varepsilon_{\bm p}^{(\nu)})-f_c^{(\nu)}(\varepsilon_{\bm p}^{(\nu)}) \right] \Phi_\nu(\bm p) \:, 
\end{equation}
\begin{equation}\label{Phi_nu}
\Phi_\nu(\bm p)  = \Biggl\{  \left[{2|v_{x,vc}^{(\nu)}|^2 \, v_{x,cc}^{(\nu)} \over \hbar\omega + {\rm i}\gamma -2 \varepsilon_{\bm p}^{(\nu)}} + {v_{x,vc}^{(\nu)}\over 2} \left({\partial v_x^{(\nu)}\over \partial p_x} \right)_{cv} \right] 
\end{equation}
\[
\times 
{ 1 \over 2\hbar \omega - 2\varepsilon_{\bm p}^{(\nu)} + {\rm i}\gamma} 
	 - {|v_{x,vc}^{(\nu)}|^2 \, v_{x,cc}^{(\nu)} \over (\hbar\omega + {\rm i}\gamma)^2-(2\varepsilon_{\bm p}^{(\nu)})^2}  \Biggr\} + c.c.(-\omega)\nonumber ,\\ 
\]
and $\varepsilon_{\bm p}^{(\nu)}$ is the electron energy in the $\nu^{\rm th}$ valley. In deriving Eqs.~\eqref{chi_nu} and~\eqref{Phi_nu} we took into account the electron-hole symmetry: $\bm v_{vv}^{(\nu)} = -\bm v_{cc}^{(\nu)}$ and $E_{cv}^{(\nu)} = 2 \varepsilon_{\bm p}^{(\nu)}$. Note that $\Phi_+(\bm p) = -\Phi_-(-\bm p)$, which indicates that $\chi_+ \neq -\chi_-$ and the second harmonic is generated only for nonequal distributions of electrons in the valleys.

We consider a valley polarized degenerate electron gas with the Fermi quasi-energies $E_{\rm F}^{(\pm)} = E_{\rm F} \pm \Delta E_{\rm F}/2$ in the $K$ and $K'$ valleys, respectively, see Fig.~\ref{fig_SHG}b. In this case, the distribution functions satisfy the condition
$f_v^{(\nu)}(-\varepsilon_{\bm p}^{(\nu)})-f_c^{(\nu)}(\varepsilon_{\bm p}^{(\nu)}) = {\theta(\varepsilon_{\bm p}^{(\nu)}-E_{\rm F}^{(\nu)})}$. For small valley polarization, when $|\Delta E_{\rm F}| \ll E_{\rm F}$, Eq.~\eqref{chi} yields
\begin{equation}\label{chi_exp}
	\chi \approx \frac{\partial \chi_+ }{\partial E_{\rm F}} \Delta E_{\rm F} =   {\rm i} \left({e \over \omega}\right)^3 \Delta E_{\rm F} \sum_{\bm p} \delta(\varepsilon_{\bm p}^{(+)}-E_{\rm F}) \Phi_+(\bm p).
\end{equation}

The trigonal warping of the electron energy spectrum in graphene responsible for SHG is small and can be considered as a perturbation.
To first order in the warping parameter $\mu$, the electron energy and the velocity matrix elements have the form
\begin{align}
&	\varepsilon_{\bm p}^{(\nu)} = v_0 p - \nu \mu p^2 \cos{3\varphi_{\bm p}} \:, \\
&	v_{x,cc}^{(\nu)} = v_0 \cos{\varphi_{\bm p}} + \nu\mu p {\cos{4\varphi_{\bm p}}-5\cos{2\varphi_{\bm p}}\over 2} \:, \\
&	v_{x,vc}^{(\nu)} = {\rm i} v_0\sin{\varphi_{\bm p}} + {\rm i} \nu\mu p {\sin{4\varphi_{\bm p}}-3\sin{2\varphi_{\bm p}} \over 2} \:, \\
& \left({\partial v_x^{(\nu)} \over \partial p_x} \right)_{cv} = 2 {\rm i} \mu \nu \sin{\varphi_{\bm p}} \:,  
\end{align}
where $\varphi_{\bm p}$ is the azimuthal angle of the $\bm p$ vector. 

Finally, summing up Eq.~\eqref{chi_exp} over the momentum we obtain
\begin{equation}\label{chi_fin}
	\chi 	= 
	{ \mu \, e^3 \hbar  \over 8\pi v_0 E_{\rm F}^2} {\Delta E_{\rm F} \over E_{\rm F}} 
	[G(\omega)+G^*(-\omega)]   \:, \quad
\end{equation}
where the complex function $G(\omega)$ is given by
\begin{align}
&G(\omega)=-{{\rm i} E_{\rm F}^4 \over (\hbar\omega)^3(\hbar\omega + {\rm i} \gamma -2E_{\rm F})} \\
&\times \left( {\hbar\omega \over \hbar\omega + {\rm i} \gamma/2 - E_{\rm F}} 
- {2E_{\rm F} \over \hbar\omega + {\rm i} \gamma+2E_{\rm F}} \right). \nonumber
\end{align}
As discussed above, the second-order susceptibility given by Eq.~\eqref{chi_fin} is proportional to the valley polarization $\Delta E_{\rm F}/E_{\rm F}$. Therefore, optical response at the double frequency can be used to measure the valley polarization in graphene. Moreover, specific polarization dependence of SHG determined by non-zero components of the tensor $\bm \chi$, see Eq.\eqref{chi_comp}, enables to discriminate the effect from possible background noise. 

\begin{figure}[t]
\includegraphics[width=0.8\linewidth]{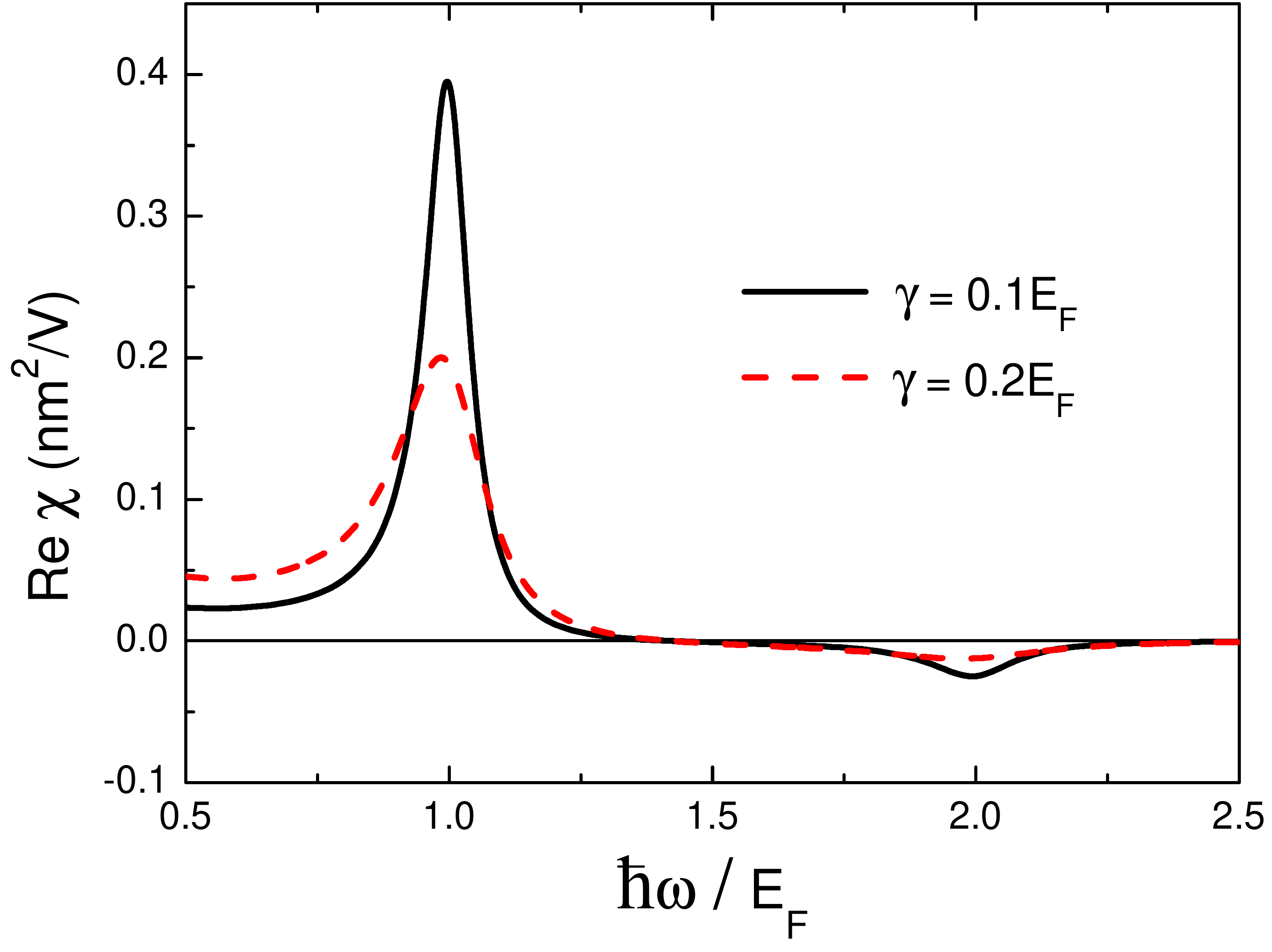}  
\includegraphics[width=0.8\linewidth]{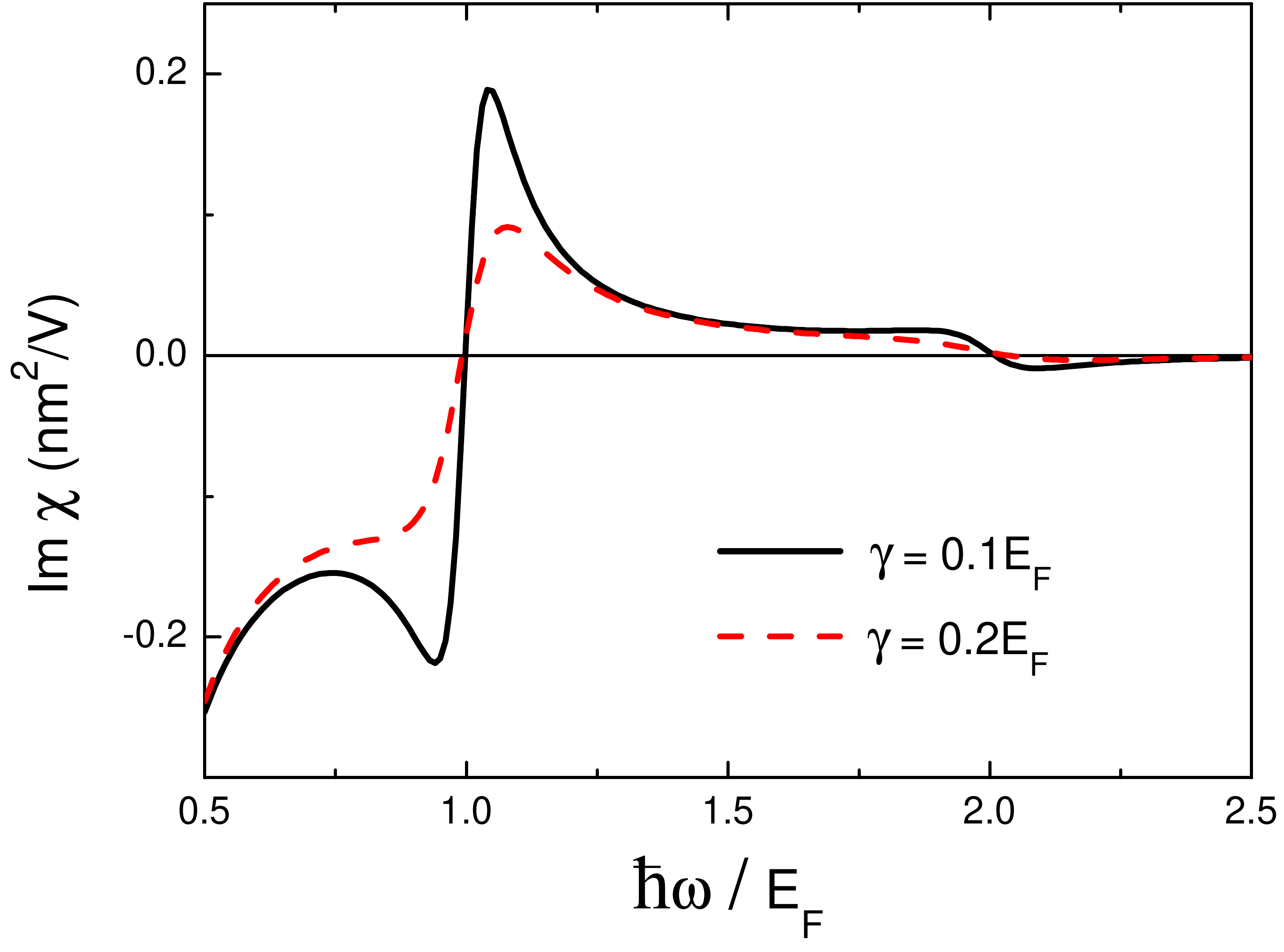}
\caption{Dependence of the real and imaginary parts of $\chi$ on the incident photon energy.   The curves are plotted after Eq.~\eqref{chi_fin} for ${\Delta E_{\rm F}/E_{\rm F}=0.1}$, $E_{\rm F}=100$~meV,  ${\mu\hbar/v_0=0.3}$~\AA, and for different decay rates $\gamma$.}
\label{fig_freq_depend}
\end{figure}

Figure~\ref{fig_freq_depend} shows the dependence of the real and imaginary parts of $\chi$ on the energy of incident photons. Valley polarization induced SHG demonstrates a resonant behavior at $\hbar \omega \approx E_{\rm F}$, which is described by  
\begin{equation}\label{chi_res}
	\chi \approx { {\rm i} \mu \, e^3 \hbar  \over 8\pi v_0 E_{\rm F} } {\Delta E_{\rm F} \over E_{\rm F}} \frac{1}{\hbar\omega - E_{\rm F} + {\rm i} \gamma/2}  \:.
\end{equation}
The resonance is situated in the spectral range where one-photon direct optical transitions are forbidden. Microscopically, it originates from a strong difference in the rates of two-photon absorption in the $K$ and $K'$ valleys due to different occupations of the final states, see Fig.~\ref{fig_SHG}b. Additional resonance at $\hbar \omega = 2 E_{\rm F}$ is situated at the edge of fundamental absorption band and stems from a difference in the one-photon absorption rates in the valleys.

The calculation yields $\chi \approx 0.4$~nm$^2$/V for the valley polarization $\Delta E_{\rm F}/E_{\rm F} =0.1$, Fermi energy $E_{\rm F}=100$~meV, photon energy $\hbar\omega=E_{\rm F}$, broadening $\gamma = 10$~meV, 
and $\mu\hbar/v_0=0.3$~\AA~\cite{CastroNeto}. 
Such a value of $\chi$ is rather high and comparable to the
nonlinear susceptibility of doped graphene induced by in-plane electric current with the density $1$~A/cm~\cite{Cheng2014}. 
We also note that nonlinear susceptibility of the same order of magnitude has been recently measured in 
MoS$_2$ and WS$_2$ monolayers, where the effect comes from the lack of crystal lattice  space inversion~\cite{MoS2_1,MoS2_2,WS2}.

To summarize, we have shown that valley polarization of free carriers in graphene can be probed by the effect of second optical harmonic generation. The effect has a specific light polarization dependence caused by the trigonal symmetry of electron valleys in graphene.

The work was supported by the Russian Foundation for Basic Research
and EU project POLAPHEN.

\end{document}